\documentclass[12pt]{amsart}
\usepackage{amssymb,amsmath}
\newtheorem{claim}{}[section]
\newtheorem{theorem}[claim]{Theorem}
\newtheorem{lemma}[claim]{Lemma}
\newtheorem{proposition}[claim]{Proposition}

\renewenvironment{proof}{\noindent{\it Proof. \hskip0pt}}
                      {$\square$\par\medskip}

\begin{document}
\baselineskip 6.2 truemm
\parindent 1.5 true pc

\title{Construction of entangled states with positive partial transposes based on indecomposable positive linear maps}
\thanks{${}^1$partially supported by a grant(R14-2003-006-01002-0) from KOSEF.}

\maketitle

\centerline{by}
\bigskip

\centerline{Kil-Chan Ha} \centerline{Department of Mathematics,
Sejong University, Seoul 143-747, KOREA} \centerline{\tt
kcha@sejong.ac.kr}
\bigskip

\bigskip
\centerline{and}
\bigskip

\centerline{Seung-Hyeok Kye${}^1$} \centerline{Department of
Mathematics, Seoul National University, Seoul 151-742, KOREA}
\centerline{\tt kye@snu.ac.kr}
\begin{abstract}
We show that every entanglement with positive partial transpose
may be constructed from an indecomposable positive linear map
between matrix algebra.
\end{abstract}

\vskip 1truecm
\baselineskip 6.2 truemm
\newcommand\lan{\langle}
\newcommand\ran{\rangle}
\newcommand\tr{{\text{\rm Tr}}\,}
\newcommand\ot{\otimes}
\newcommand\wt{\widetilde}
\newcommand\join{\vee}
\newcommand\meet{\wedge}
\renewcommand\ker{{\text{\rm Ker}}\,}
\newcommand\im{{\text{\rm Im}}\,}
\newcommand\mc{\mathcal}
\newcommand\transpose{{\text{\rm t}}}
\newcommand\FP{{\mathcal F}({\mathcal P}_n)}
\newcommand\ol{\overline}
\newcommand\JF{{\mathcal J}_{\mathcal F}}
\newcommand\FPtwo{{\mathcal F}({\mathcal P}_2)}
\newcommand\hada{\circledcirc}
\newcommand\id{{\text{\rm id}}}
\newcommand\tp{{\text{\rm tp}}}
\newcommand\pr{\prime}
\newcommand\inte{{\text{\rm int}}\,}
\newcommand\ttt{{\text{\rm t}}}
\newcommand\spa{{\text{\rm span}}\,}
\newcommand\conv{{\text{\rm conv}}\,}
\newcommand\rank{\ {\text{\rm rank}}\,}
\newcommand\vvv{\mathbb V_{m\meet n}\cap\mathbb V^{m\meet n}}
\newcommand\re{{\text{\rm Re}}\,}
\newcommand\la{\lambda}
\newcommand\msp{\hskip 2pt}
\newcommand\ppt{\mathbb T}
\newcommand\cc{\mathbb{C} }
\newcommand\pp{\mathbb{P} }
\newcommand\rr{\mathbb{R} }
\newcommand\range{{\mathcal R}}

\section{Introduction}

The theory of entanglement in quantum physics is now playing a key
r\^ ole in the quantum information theory and quantum
communication theory. In order to characterize various kinds of
entangled states, the notion of positive linear maps is turned out to be
very useful as was developed by quantum physicists \cite{hor96}, \cite{hor97} and \cite{terhal00}.
On the other hand, some operator algebraists \cite{eom-kye}, \cite{stormer82}, \cite{woronowicz}
studied various kinds of positive linear maps
and showed their non-decomposability using several kinds of block matrices, which are
nothing but entanglement.

The duality between entangled states and positive linear maps is also used to construct them.
The notion of entanglement was used \cite{terhal01} to
construct a new kind of indecomposable positive linear maps. This construction
was generalized in \cite{lew00}, \cite{lew01} to give a method of creating
entanglement witnesses and corresponding positive linear maps
starting from the edge states. Thus they presented a
canonical form of non-decomposable entanglement witnesses and the
corresponding indecomposable positive linear maps.

On the other hand, the authors \cite{ha+kye} have
constructed  entangled states  with positive partial transposes
from decomposable positive linear maps. This methods provided us
an explicit example of one parameter family of
$3\otimes 3$ edge states
which are not arising from unextendible product basis.

In the present Letter, we show that every edge state with positive
partial transpose arises from an indecomposable positive linear
map in the way described in the previous paper \cite{ha+kye}.
Therefore, our result complements the opposite direction of
\cite{lew00} and \cite{lew01}, where it was shown that, given an
edge state, its entanglement witness (and so the corresponding
linear map) can be determined. For the interplay between the
notions of entanglement and positive linear maps, we refer to the
paper \cite{ha+kye} together with the references there.

As in the paper \cite{ha+kye}, we denote by $\mathbb D$ the convex cone
of all decomposable positive linear maps from the $C^*$-algebra
$M_m$ of all $m\times m$ matrices over the complex field into
$M_n$, and by $\mathbb T$ the cone of all positive semi-definite
$m\times m$ matrices over $M_n$ whose partial transposes are also
positive semi-definite. The key step is to show that every face of
the cone $\mathbb T$ is exposed. This is quite surprising, since not
every face of the dual cone $\mathbb D$ of $\mathbb T$ is exposed
as was shown in \cite{byeon-kye} and \cite{kye-decom}.

We first develop in Section 2 some generalities about duality
theory between the cones $\mathbb D$ and $\mathbb T$, and
characterize faces of the cone $\mathbb T$. In Section 3, we show
that every face of the cone $\mathbb T$ is exposed, and find all
faces of $\mathbb T$ in the simplest nontrivial case of $m=n=2$.
In the final section, we show that edge state with
positive partial transpose arises from an indecomposable positive
linear map in the way described in \cite{ha+kye}. Using our
terminologies, it is easy to check the following statement: An
entangled state $\rho$ with positive partial transpose is an edge state if and only if the proper
face of $\mathbb T$ containing $\rho$ as an interior point does not
contain a separable state.

Throughout this Letter, we will not use bra-ket notation as in the
previous paper \cite{ha+kye}. Every vector will be considered as a
column vector. If $x\in\mathbb C^m$ and $y\in \mathbb C^n$ then
$x$ will be considered as an $m\times 1$ matrix, and $y^*$ will be
considered as a $1\times n$ matrix, and so $xy^*$ is an $m\times
n$ rank one matrix whose range is generated by $x$ and whose
kernel is orthogonal to $y$. For a vector $x$, the notation $\ol
x$ will be used for the vector whose entries are conjugate of the
corresponding entries. The notation $\lan\,\cdot\, ,\cdot\, \ran$
will be used for bi-linear pairing. On the other hand, $(\,\cdot\,
|\,\,\cdot\, )$ will be used for the inner product, which is
sesqui-linear, that is, linear in the first variable and
conjugate-linear in the second variable.

The authors are grateful to the referee of this Letter for
bringing their attentions to the paper \cite{lew01}

\section{Duality between two cones}

In this section, we first consider the faces of the cones
$\conv(C_1,C_2)$ and $C_1\cap C_2$ of the cones $C_1$ and $C_2$,
where $\conv(C_1,C_2)$ denotes the convex hull generated by $C_1$
and $C_2$. We have already seen in \cite{kye-decom} that every
face $F$ of the cone $C=\conv(C_1,C_2)$ associates with a unique
pair $(F_1,F_2)$ of faces of $C_1$ and $C_2$, respectively, with
the properties
\begin{equation}\label{join-face}
F=\conv(F_1,F_2),\qquad F_1=F\cap C_1,\qquad F_2=F\cap C_2.
\end{equation}
Actually, it is easy to see that $F_i=F\cap C_i$ is a face of
$C_i$ for $i=1,2$ and the identity $F=\conv(F_1,F_2)$ holds. It
should be noted \cite{byeon-kye} that
$\conv(F_1,F_2)=\conv(E_1,F_2)$ may be a face of $C$ even though
$E_1\neq F_1$. If $F_1$ and $F_2$ satisfy the conditions in
(\ref{join-face}) then we will write
$$
\sigma(F_1,F_2):=\conv(F_1,F_2).
$$
We always assume that
\begin{equation}\label{ass-join}
\sigma(F_1,F_2)\cap C_i=F_i,\qquad i=1,2,
\end{equation}
whenever we use the notation $\sigma(F_1,F_2)$.

On the other hand, if $F_i$ is a face of the cone $C_i$ for $i=1,2$ then $F_1\cap F_2$ is a face of $C_1\cap C_2$. Conversely,
every face $F$ of the cone $C=C_1\cap C_2$ associates with a unique pair $(F_1,F_2)$ of faces of $C_1$ and $C_2$, respectively,
with the properties
$$
F=F_1\cap F_2,\qquad \inte F\subset\inte F_1,\qquad \inte
F\subset\inte F_2,
$$
where $\inte F$ denotes the set of all interior point of the convex set $F$, which is the topological
interior of $F$ relative to the hyperplane generated by $F$.
To see that, take an interior point $x$ of $F$ in $C_1\cap C_2$. If we take the face $F_i$ of $C_i$ with $x\in\inte F_i$ for
$i=1,2$ then we have
$$
x\in\inte F_1\cap\inte F_2\subset\inte(F_1\cap F_2).
$$
Since $F_1\cap F_2$ is a face of $C$, we conclude that $F=F_1\cap F_2$. The uniqueness is clear, because every convex set is
decomposed into the interiors of faces. We will use the notation
$$
\tau(F_1,F_2):=F_1\cap F_2
$$
only when
\begin{equation}\label{ass-cap}
\inte \tau(F_1,F_2)\subset\inte F_i,\qquad i=1,2.
\end{equation}

Now, we proceed to consider the duality.
Let $X$ and $Y$ be finite dimensional normed spaces, which are dual each other
with respect to a bilinear pairing $\lan\ , \ \ran$.
For a subset $C$ of $X$ (respectively $D$ of $Y$), we
define the {\sl dual cone} $C^\circ$ (respectively $D^\circ$) by the set of all
$y\in Y$ (respectively $x\in X$) such that $\lan x,y\ran \ge 0$ for each $x\in
C$ (respectively $y\in D$). It is clear that $C^{\circ\circ}$ is the closed
convex cone of $X$ generated by $C$.
We have
$$
[\conv(C_1,C_2)]^\circ=C_1^\circ\cap C_2^\circ,\qquad (C_1\cap
C_2)^\circ =\conv(C_1^\circ, C_2^\circ),
$$
whenever $C_1$ and $C_2$ are closed convex cones of $X$, as was seen in \cite{eom-kye}.
For a subset $S$ of a closed convex cone $C$ of $X$, we define the subset $S^\pr$
of $C^\circ$ by
$$
S^\pr=\{ y\in C^\circ:\lan x,y\ran =0\ {\text{\rm for each}}\ x\in S\}.
$$
It is then clear that $S^\pr$ is a closed face of $C^\circ$. If we
take an interior point $x_0$ of a face $F$ then we see that
$F^\prime=\{x_0\}^\prime$. We will write $x_0^\prime$ for $\{x_0\}^\prime$.

Let $F_i$ be a face of the convex cone $C_i$, for $i=1,2$,
satisfying the conditions in (\ref{join-face}) then
$\sigma(F_1,F_2)=\conv(F_1,F_2)$ is a face of $C=\conv(C_1,C_2)$.
It is easy to see that
\begin{equation}\label{dual-basic}
\sigma(F_1,F_2)^\prime=F_1^\prime\cap F_2^\prime,
\end{equation}
where it should be noted that the dual faces should be taken in
the corresponding duality. For example, $\sigma(F_1,F_2)^\prime$
is the set of all $y\in C^\circ=C_1^\circ\cap C_2^\circ$ such that
$\lan x,y\ran =0$ for each $x\in\conv(F_1,F_2)$. On the other
hand, $F_i^\prime$ is the set of all $y\in C_i^\circ$ such that
$\lan x,y\ran=0$ for each $x\in F_i$. Analogously, we also have
\begin{equation}\label{dual-basic-1}
\tau(F_1,F_2)^\prime= \conv(F_1^\prime,F_2^\prime).
\end{equation}
From the easy inclusion $F_i^\prime\subset\tau(F_1,F_2)^\prime$,
one direction comes out. For the reverse inclusion, let $y\in
\tau(F_1,F_2)^\prime$. Since $y\in (C_1\cap C_2)^\circ=
\conv(C_1^\circ,C_2^\circ)$, we may write $y=y_1+y_2$ with $y_i\in
C_i^\circ$ for $i=1,2$. We also take an interior point $x$ of
$\tau(F_1,F_2)$. Then we have $x\in\inte F_i\subset C_i$ by
(\ref{ass-cap}), and so $\lan x,y_i\ran\ge 0$ for $i=1,2$. From
the relation
$$
0=\lan x,y\ran=\lan x,y_1\ran+\lan x,y_2\ran,
$$
we conclude that $\lan x,y_i\ran=0$. Since $x$ is an interior
point of $F_i$, we see that $y_i\in F_i^\prime$ for $i=1,2$, and
$y\in\conv(F_1^\prime,F_2^\prime)$.

Now, we consider the cone $C_1=\mathbb P_{m\meet n}$ of all completely positive linear maps from $M_m$ into $M_n$, and
the cone $C_2=\mathbb P^{m\meet n}$ of all completely copositive linear maps from $M_m$\ into $M_n$. Both cones
are sitting in the linear space ${\mathcal L}(M_m,M_n)$ of all linear maps from $M_m$ into $M_n$.
It is well-known that every map in the cone $\mathbb P_{m\meet n}$
(respectively $\mathbb P^{m\meet n}$) is of the form
$\phi_\mathcal V$ (respectively $\phi^\mathcal V$)
for a subset ${\mathcal V}=\{V_1, V_2,\dots, V_\nu\}$ of
$M_{m\times n}$:
$$
\begin{aligned}
&\phi_\mathcal V: X\mapsto \sum_{i=1}^\nu V_i^*XV_i,\qquad X\in
M_m,\\
&\phi^\mathcal V: X\mapsto \sum_{i=1}^\nu V_i^*X^{\ttt}V_i,\qquad
X\in M_m,
\end{aligned}
$$
where $X^\ttt$ denote the transpose of $X$.
We denote by $\phi_V=\phi_{\{V\}}$ and $\phi^V=\phi^{\{V\}}$.
For a subspace $E$ of $M_{m\times n}$, we define
$$
\begin{aligned}
\Phi_E=&\{\phi_\mathcal V\in\mathbb P_{m\meet n}: \spa{\mathcal
V}\subset E\},\\
\Phi^E=&\{\phi^\mathcal V\in\mathbb P^{m\meet n}: \spa{\mathcal
V}\subset E\},
\end{aligned}
$$
where $\spa{\mathcal V}$ denotes the span of the set $\mathcal V$.
We have shown in \cite{kye-cambridge} that the correspondence
$$
E\mapsto \Phi_E\qquad {\text{\rm (respectively}}\
E\mapsto \Phi^E{\text{\rm )}}
$$
gives rise to a lattice isomorphism between the lattice
of all subspaces of the vector space $M_{m\times n}$ and the lattice
of all faces of the convex set $\mathbb P_{m\meet n}$ (respectively $\mathbb P^{m\meet n}$). It is also known that
$$
\inte \Phi_E=\{\phi_\mathcal V\in\mathbb P_{m\meet n}: \spa{\mathcal
V}= E\}
$$
and similarly for $\inte \Phi^E$.

A linear map in the cone
$$
\mathbb D:= \conv (\mathbb P_{m\meet n}, \mathbb P^{m\meet
n})\subset {\mathcal L}(M_m,M_n)
$$
is said to be {\sl decomposable}. Every decomposable map is
positive, that is, sends positive semi-definite matrices into
themselves, but the converse is not true. There are bunch of
examples of indecomposable positive linear maps in the literature
as was referred in \cite{ha+kye}. We see that every face of the
cone $\mathbb D$ is of the form $\conv (\Phi_D,\Phi^E)$ for a pair
$(D,E)$ of subspaces of $M_{m\times n}$. We say that a pair
$(D,E)$ is a {\sl decomposition pair} if the set $\conv
(\Phi_D,\Phi^E)$ is a face of $\mathbb D$ together with the
corresponding condition (\ref{ass-join}). By an abuse of notation,
we will write
$$
\sigma (D,E):=\sigma(\Phi_D,\Phi^E)
$$
for a decomposition pair $(D,E)$. The condition (\ref{ass-join}) is then written by
$$
\sigma(D,E)\cap \mathbb P_{m\meet n}= \Phi_D,\qquad
\sigma(D,E)\cap \mathbb P^{m\meet n}= \Phi^E.
$$
In this way, every decomposition pair gives rise to a face of $\mathbb D$, and every face of $\mathbb D$ corresponds
to a unique decomposition pair.

In \cite{eom-kye}, we have considered the bi-linear pairing between
the spaces ${\mathcal L}(M_m, M_n)$ and $M_n\otimes M_m$, given by
\begin{equation}\label{definition}
\lan A,\phi\ran =\tr \left[ \left(\sum_{i,j=1}^m
\phi(e_{ij}) \otimes e_{ij}\right) A^\ttt \right]
=\sum_{i,j=1}^m\lan\phi(e_{ij}),a_{ij}\ran,
\end{equation}
for $A=\sum_{i,j=1}^m a_{ij}\ot e_{ij}\in M_n\ot M_m$ and
$\phi\in{\mathcal L}(M_m, M_n)$, where the
bi-linear form in the right-side is given by $\lan X,Y\ran=\tr (YX^\ttt )$
for $X,Y\in M_n$. It is well-known \cite{choi75-10} (see also \cite{eom-kye}) that
the dual cone $(\mathbb P_{m\meet n})^\circ=(M_n\otimes M_m)^+$ consists of all
positive semi-definite block matrices in $M_n\otimes M_m$.
Recall that every element in the tensor product $M_n\otimes M_m$ may be considered as an $m\times m$ matrix over $M_n$.
We define the {\sl partial transpose} or {\sl block transpose} $A^\tau$ by
$$
\left(\sum_{i,j=1}^m a_{ij}\otimes
e_{ij}\right)^\tau=\sum_{i,j=1}^m a_{ji}\otimes e_{ij}.
$$
Then we also see that the dual cone $(\mathbb P^{m\meet n})^\circ$ consists of all block matrices in $M_n\otimes M_m$
whose block transposes are positive semi-definite.

We identify a matrix $z\in M_{m\times n}$  and a vector
$\wt z\in \mathbb C^n\otimes\mathbb C^m$ as follows: For $z=[z_{ik}]\in
M_{m\times n}$, define
$$
\begin{aligned}
z_i=&\sum_{k=1}^n z_{ik} e_k\in \mathbb C^n,\qquad i=1,2,\dots, m,\\
\wt z=&\sum_{i=1}^m z_i\otimes e_i\in \mathbb C^n\otimes\mathbb C^m.
\end{aligned}
$$
Then $z\mapsto \wt z$ defines an inner product isomorphism
from $M_{m\times n}$ onto $\mathbb C^n\otimes \mathbb C^m$. We also note that
$\wt z\,{\wt z}^*$ is a positive semi-definite matrix in $M_n\otimes M_m$ of rank one, and
the cone $(\mathbb P_{m\meet n})^\circ$ is spanned by $\{\wt z\,{\wt z}^*: z\in M_{m\times n}\}$.
We have shown in \cite{kye-decom} that
\begin{equation}\label{eq1}
\lan \wt z\,{\wt z}^*, \phi_V\ran =
\lan (\wt z\,{\wt z}^*)^\tau, \phi^V\ran
=|(z|V)|^2,
\end{equation}
for $m\times n$ matrices $z,V\in M_{m\times n}$. Therefore, we see that
$\wt z\,{\wt z}^*$ belongs to $(\Phi_D)^\prime$ if and only if $z\in D^\perp$, and so we have
$$
(\Phi_D)^\prime=\conv \{ \wt z\,{\wt z}^*: z\in D^\perp\}\\
=\{A\in (M_n\otimes M_m)^+: {\mathcal R}A\subset {\wt D}^\perp\},
$$
where ${\mathcal R}A$ is the range space of $A$ and $\wt D=\{\wt z\in\mathbb C^n\otimes\mathbb C^m: z\in D\}$. If we denote by
$$
\begin{aligned}
\Psi_D&=\{A\in (M_n\otimes M_m)^+: {\mathcal R}A\subset \wt D\},\\
\Psi^E&=\{A\in M_n\otimes M_m:  A^\tau\in\Psi_E\}
\end{aligned}
$$
then we have
\begin{equation}\label{nnnn}
\inte\Psi_D=\{A\in \Psi_D: {\mathcal R}A= {\wt D}\},\qquad
\inte\Psi^E=\{A\in \Psi^E: {\mathcal R}A^\tau=\wt E\}.
\end{equation}
In short, we have
$$
(\Phi_D)^\prime =\Psi_{D^\perp},\qquad
(\Psi_D)^\prime =\Phi_{D^\perp},\qquad
(\Phi^E)^\prime =\Psi^{E^\perp},\qquad
(\Psi^E)^\prime =\Phi^{E^\perp}.
$$
It is well known that every face of the cone $(M_n\otimes M_m)^+$ is of the form $\Psi_D$ for a subspace $D$ of
$M_{m\times n}$.

We say that a pair $(D,E)$ of subspaces of $M_{m\times n}$ is an {\sl intersection pair} if the faces
$\Psi_D$ and $\Psi^E$ satisfy the condition (\ref{ass-cap}). More precisely, a pair $(D,E)$ is an intersection pair
if and only if
$$
\inte (\Psi_D\cap \Psi^E)\subset \inte\Psi_D\cap \inte\Psi^E.
$$
Then, every face of the cone
$$
\mathbb T:=(\mathbb P_{m\meet n})^\circ\cap(\mathbb P^{m\meet n})^\circ
=\{A\in (M_n\otimes M_m)^+: A^\tau\in(M_n\otimes M_m)^+\}
$$
is of the form $\tau(\Psi_D,\Psi^E)$ for a unique intersection pair $(D,E)$. We also write
$$
\tau(D,E):=\tau(\Psi_D,\Psi^E)
$$
if $(D,E)$ is an intersection pair,
by an another abuse of notation. By (\ref{dual-basic}) and (\ref{dual-basic-1}), we have
\begin{equation}\label{bbbb}
\sigma(D,E)^\prime=\Psi_{D^\perp}\cap\Psi^{E^\perp},\qquad
\tau(D,E)^\prime=\conv (\Phi_{D^\perp},\Phi^{E^\perp}).
\end{equation}
It should be noted that $(D^\perp, E^\perp)$ need not to be an intersection pair even if $(D,E)$ is a decomposition pair.

We denote by $\mathbb P_1$ the cone of all positive linear maps from $M_m$ into $M_n$.
Then the dual cone $\mathbb V_1:=(\mathbb P_1)^\circ$ of the cone
$\mathbb P_1$ with respect to the pairing (\ref{definition}) is given by
$$
\mathbb V_1=\conv\{\wt z\,\wt z^*: {\text{\rm rank}}\ z=1\}.
$$
By the relation
$$
\wt{xy^*}\,{\wt{xy^*}}^*=(\ol y\otimes x)(\ol y\otimes x)^*
=\ol y\,{\ol y}^*\otimes xx^*,
$$
we have $\mathbb V_1=M_n^+\otimes M_m^+$. It is easy to see that $\mathbb V_1\subset \mathbb T$.
A density matrix in $\mathbb T\setminus\mathbb V_1$ is said to be an {\sl entangled state with positive partial transpose}.

\section{Exposed faces}

A face $F$ of a convex set $C$ is said to be {\sl exposed} if there is $x\in C^\circ$ such that
$F=x^\prime$. It is easy to see that a face $F$ is exposed if and only if $F=F^{\prime\prime}$. A decomposition pair
(respectively an intersection pair)
$(D,E)$ is said to be {\sl exposed} if the face $\sigma(D,E)$ (respectively $\tau(D,E)$) is exposed.

\begin{lemma}\label{lemma-1}
Let $A\in\mathbb T$. If $A^\prime=\sigma(D,E)$ then ${\mathcal R}A={\wt D}^\perp$ and ${\mathcal R}A^\tau={\wt E}^\perp$.
\end{lemma}

\begin{proof}
First of all, the relation
$$
A\in A^{\prime\prime}=\sigma(D,E)^\prime=\Psi_{D^\perp}\cap\Psi^{E^\perp}
$$
implies that ${\mathcal R}A\subset{\wt D}^\perp$ and ${\mathcal R}A^\tau\subset{\wt E}^\perp$.
For the reverse inclusion,
let $V\in M_{m\times n}$ with $\wt V\in ({\mathcal R}A)^\perp$, and write
$A=\sum_i \wt{z_i}{\wt{z_i}}^*$ with $z_i\in M_{m\times n}$.
Then we have
$$
\lan A,\phi_V\ran =\sum |(z_i\, |\, V)|^2=0,
$$
and $\phi_V\in A^\prime$. Since
$A^\prime=\sigma(D,E)$ by the
assumption, we have
$$
\phi_V\in A^\prime\cap \mathbb P_{m\meet n}=\sigma(D,E)\cap\mathbb P_{m\meet n}=\Phi_D.
$$
This implies $V\in D$, and so we have ${\mathcal R}A={\wt D}^\perp$.
For the second relation
${\mathcal R}A^\tau={\wt E}^\perp$,
we note the following easy identities
$$
\lan A^\tau,\phi_V\ran =\lan A,\phi^V\ran,\qquad \lan A^\tau,\phi^W\ran=\lan A,\phi_W\ran,
$$
by (\ref{eq1}).
This implies that $A^\prime=\sigma(D,E)$ if and only if $(A^\tau)^\prime=\sigma(E,D)$.
Therefore, the second relation ${\mathcal R}A^\tau={\wt E}^\perp$ follows from the first.
\end{proof}

\begin{proposition}\label{3333}
Let $(D,E)$ be a pair of subspaces of $m\times n$ matrices. Then
$(D,E)$ is an exposed decomposition pair if and only if $(D^\perp,
E^\perp)$ is an intersection pair.
\end{proposition}

\begin{proof}
Assume that $(D,E)$ is an exposed decomposition pair, and
take an element $A\in\inte \sigma(D,E)^\prime$. Then we have
$$
A^\prime=\sigma(D,E)^{\prime\prime}=\sigma(D,E)
$$
by the assumption. This implies that ${\mathcal R}A={\wt D}^\perp$ and ${\mathcal R}A^\tau={\wt E}^\perp$ by Lemma \ref{lemma-1},
and so we see that $A\in\inte\Psi_{D^\perp}\cap \inte\Psi^{E^\perp}$ by (\ref{nnnn}). This proves the required relation
$$
\inte (\Psi_{D^\perp}\cap \Psi^{E^\perp})=\inte\sigma(D,E)^\prime\subset \inte\Psi_{D^\perp}\cap \inte\Psi^{E^\perp}
$$
by the relation (\ref{bbbb}). Therefore, we see that $(D^\perp, E^\perp)$ is an intersection pair.

For the converse, assume that $(D^\perp,E^\perp)$ is an
intersection pair. First of all, we see that $\conv
(\Phi_{D},\Phi^{E})=\tau(D^\perp,E^\perp)^\prime$ is an exposed
face of $\mathbb D$. We may take  an exposed decomposition pair
$(D_1,E_1)$ such that $\conv (\Phi_{D},\Phi^{E})=\sigma(D_1,E_1)$.
It suffices to show that $D=D_1$ and $E=E_1$. To do this, take
$A\in\inte\tau(D^\perp, E^\perp)$. Then we have
$A\in\inte\Psi_{D^\perp}\cap \inte\Psi^{E^\perp}$ since $(D^\perp,
E^\perp)$ is an intersection pair, and so
$$
{\wt D}^\perp ={\mathcal R}A,\qquad
{\wt E}^\perp ={\mathcal R}A^\tau,\qquad
$$
by (\ref{nnnn}). On the other hand, we also have $A^\prime=\tau(D^\perp, E^\perp)^\prime=\sigma(D_1,E_1)$, and
$$
{\wt D_1}^\perp ={\mathcal R}A,\qquad
{\wt E_1}^\perp ={\mathcal R}A^\tau,\qquad
$$
by Lemma \ref{lemma-1}. Therefore, we have $D=D_1$ and $E=E_1$.
\end{proof}

\begin{theorem}\label{theorem}
Every face of the convex cone $\mathbb T$ consisting of all positive semi-definite block matrices with positive
semi-definite block transpose is exposed.
\end{theorem}

\begin{proof}
Every face of $\mathbb T$ is of the form $\tau(D,E)$ for an intersection pair $(D,E)$ of spaces of matrices.
Then $(D^\perp,E^\perp)$ is an exposed decomposition pair by Proposition \ref{3333},
and so $\tau(D,E)^\prime=\sigma(D^\perp,E^\perp)$. Therefore, we have
$$
\tau(D, E)^{\prime\prime}=\sigma(D^\perp,E^\perp)^\prime=\tau(D,E)
$$
by (\ref{bbbb}).
\end{proof}

For the simplest case of $m=n=2$, every decomposition pair has been
characterized in \cite{byeon-kye}. See also \cite{kye-2by2_II}.
For every $x,y\in\mathbb C^2$ the pair
\begin{equation}\label{class-1}
([xy^*], [\ol xy^*])
\end{equation}
is an intersection pair, where $[z_i]$ denotes the
span of $\{z_i\}$. These give us all extremal rays of the cone $\mathbb T$.
We note that $\wt {xy^*}=\ol y\otimes x$, and so the face $\tau([xy^*],[\ol xy^*])$
consists of nonnegative scalar multiples of
$(\ol y\otimes x)(\ol y\otimes x)^*
=\ol y\,{\ol y}^*\otimes xx^*$.
For every vectors $x,y,z,w$, the pair
\begin{equation}\label{class-2}
([xy^*, zw^*], [\ol xy^*, \ol zw^*])
\end{equation}
is also an exposition pair, where $x\nparallel z$ or $y\nparallel w$. Here, $x\nparallel z$ means that $x$ and $z$ are not
parallel each other.

Next, we also have intersection pairs whose subspaces spanned by three rank one matrices. They are
\begin{equation}\label{class-6}
([V]^\perp, [W]^\perp)
\end{equation}
with rank two matrices $V$ and $W$ satisfying the properties in Proposition 3.6 of \cite{byeon-kye}, or
\begin{equation}\label{class-7}
([xy^*]^\perp, [\ol xy^*]^\perp)
\end{equation}
for arbitrary vectors $x,y\in \mathbb C^2$.
For the case (\ref{class-6}), note that if $([V], [W])$ is an exposed decomposition pair then we may take vectors
$x_i,y_i$, $i=1,2,3$, such that
$$
Vy_i\perp x_i,\qquad i=1,2,3,
$$
and we have
$$
[V]^\perp=[x_1y_1^*, x_2y_2^*,x_3y_3^*],\qquad
[W]^\perp=[\ol x_1y_1^*, \ol x_2y_2^*, \ol x_3y_3^*]
$$
as is in the proof of \cite{byeon-kye} Proposition 3.6.
In the case (\ref{class-7}), we may take vector $z,w$ with $z\perp x$ and $w\perp y$
so that
$$
[xy^*]^\perp=[zy^*, xw^*, zw^*], \qquad [\ol xy^*]^\perp=[\ol zy^*, \ol xw^*, \ol zw^*].
$$
The case (\ref{class-7}) gives rise to a maximal face of the cone $\mathbb T$.

Finally, we also have intersection pairs
\begin{equation}\label{class-8}
([V]^\perp, M_2)
\end{equation}
and
\begin{equation}\label{class-9}
(M_2, [W]^\perp)
\end{equation}
with rank two matrices $V$ and $W$. These also give us maximal faces.
We may take four rank one matrices $x_iy_i^*$, $i=1,2,3,4,$ so that
$$
[V]^\perp=[x_iy_i^*: i=1,2,3,4],\qquad M_2=[\ol x_iy_i^*: i=1,2,3,4]
$$
by Proposition 3.7 of \cite{byeon-kye}. For example, consider the pair $([I]^\perp, M_2)$, where $I$ is the identity matrix.
If we take
$$
\left(\begin{matrix}1\\0\end{matrix}\right)\quad
\left(\begin{matrix}0\\1\end{matrix}\right)\quad
\left(\begin{matrix}0\\1\end{matrix}\right)\quad
\left(\begin{matrix}1\\0\end{matrix}\right)\quad
\left(\begin{matrix}1\\-1\end{matrix}\right)\quad
\left(\begin{matrix}1\\1\end{matrix}\right)\quad
\left(\begin{matrix}-i\\1\end{matrix}\right)\quad
\left(\begin{matrix}1\\-i\end{matrix}\right)
$$
for $x_1,y_1,x_2,y_2,x_3,y_3,x_4,y_4$, then we have
$$
\sum_{i=1}^4\wt{x_iy_i^*} {\wt{x_iy_i^*}}^*
=\left(
\begin{matrix}
2&1-i&-1-i&-2\\
1+i&3&0&-1-i\\
-1+i&0&3&1-i\\
-2&-1+i&1+i&2
\end{matrix}
\right)
$$
whose range space is ${\wt I}^\perp$, which is $3$-dimensional. On
the other hand, $\sum_i\wt{\ol x_iy_i^*} {\wt{\ol x_iy_i^*}}^*$ is
the block transpose of the above matrix, and the range space is
$4$-dimensional. These six cases (\ref{class-1})-(\ref{class-9})
list up all faces of the convex cone $\mathbb T$ for the case of
$m=n=2$.

\section{Entangled states with positive partial transposes}

In order to construct entanglement with positive partial
transpose, we have shown in \cite{ha+kye} that if $(D,E)$ is a
decomposition pair so that $\sigma(D,E)$ is a proper face of
$\mathbb D$ such that $\inte\sigma(D,E)\subset\inte\mathbb P_1$
then every non-zero element $A$ of the face $\sigma(D,E)^\prime$
belongs to $\mathbb T\setminus \mathbb V_1$. To begin with, we
show the converse. Note that we have the following two cases
$$
\inte\sigma(D,E)\subset\inte\mathbb P_1\qquad
{\text{\rm or}}\qquad
\sigma(D,E)\subset\partial\mathbb P_1,
$$
since $\sigma(D,E)$ is a convex subset of the cone $\mathbb P_1$, where
$\partial C:=C\setminus \inte C$ denotes the boundary of the convex set $C$.

\begin{theorem}\label{theorem2}
Let $(D,E)$ be a decomposition pair which gives rise to a proper face of $\mathbb D$. Then
the following are equivalent:
\begin{enumerate}
\item[(i)]
$\inte\sigma(D,E)\subset\inte\mathbb P_1$,
\item[(ii)]
$\sigma(D,E)^\prime\setminus\{0\}\subset \mathbb T\setminus\mathbb V_1$.
\end{enumerate}
\end{theorem}

\begin{proof}
The direction (i) $\Longrightarrow$ (ii) was proved in Theorem 1 of \cite{ha+kye}. For the reverse direction,
it suffices to show that
$$
\sigma(D,E)\subset \partial\mathbb P_1 \ \Longrightarrow
\sigma(D,E)^\prime\cap \mathbb V_1 \supsetneqq \{0\}.
$$
To do this, assume that $\sigma(D,E)\subset \partial\mathbb P_1$. Take $\phi\in\inte \sigma(D,E)$, and take the face
$F$ of $\mathbb P_1$ such that $\phi\in \inte F$. We note that $F$ is a proper face of $\mathbb P_1$ since $\phi\in\partial\mathbb P_1$
by the assumption. We also note that $F$ is a face of $\mathbb P_1=(\mathbb V_1)^\circ$ and $\sigma(D,E)$ is a face of
$\mathbb D=\mathbb T^\circ$, and so we have
$$
\begin{aligned}
F^\prime&=\{A\in\mathbb V_1: \langle A,\phi\rangle =0\}\\
\sigma(D,E)^\prime&=\{A\in\mathbb T: \langle A,\phi\rangle=0\}.
\end{aligned}
$$
This shows that $\sigma(D,E)^\prime\cap \mathbb V_1=F^\prime$,
which has a non-zero element since $F$ is a proper face of
$\mathbb P_1$.
\end{proof}

Now, we show that every edge state $A$ with positive partial transpose arise in the way described in \cite{ha+kye}.
First of all, we take the face $F$ of $\mathbb T$ such that $A\in\inte F$. Since $A$ is an edge state, we see that
$F$ has no separable state.
We also see that $F=\sigma(D,E)^\prime$ for an exposed decomposition pair $(D,E)$ by Theorem \ref{theorem}, and so we have
$$
\sigma(D,E)^\prime\setminus\{0\} = F\setminus\{0\}\subset \mathbb T\setminus\mathbb V_1.
$$
It follows from Theorem \ref{theorem2} that
\begin{equation}\label{sssdfg}
\inte\sigma(D,E)\subset\inte\mathbb P_1.
\end{equation}
Of course, a face $\sigma(D,E)$ of $\mathbb D$ with the property (\ref{sssdfg}) arises
from an indecomposable positive linear map in $\mathbb P_1\setminus\mathbb D$, as was explained in \cite{ha+kye}.


\begin{thebibliography}{99}
\bibitem{byeon-kye} E.-S. Byeon and S.-H. Kye, \it Facial
structures for positive linear maps in the two dimensional matrix
algebra, \rm Positivity, \bf 6 \rm (2002), 369--380.

\bibitem{choi75-10}  M.-D. Choi, \it Completely positive linear
maps on complex matrices, \rm Linear Alg. Appl. \bf 10 \rm (1975),
285--290.

\bibitem{eom-kye} M.-H. Eom and S.-H. Kye, \it Duality for positive linear maps in matrix
algebras, \rm Math. Scand. \bf 86 \rm (2000), 130--142.

\bibitem{ha+kye}
K.-C. Ha, S.-H. Kye and Y. S. Park, \it Entanglements with
positive partial transposes arising from indecomposable positive
linear maps, \rm Phys. Lett. A \bf 313 \rm (2003), 163--174.


\bibitem{hor96} M. Horodecki, P. Horodecki and R. Horodecki, \it
Separability of mixed states: necessary and sufficient conditions,
\rm Phys. Lett. A \bf 223 \rm (1996), 1--8.

\bibitem{hor97} P. Horodecki, \it Separability criterion and
inseparable mixed states with positive partial transposition, \rm
Phys. Lett. A \bf 232 \rm (1997), 333--339.

\bibitem{kye-cambridge} S.-H. Kye, \it On the convex
set of all completely positive linear maps in matrix algebras, \rm
Math. Proc. Cambridge Philos. Soc. \bf 122 \rm (1997), 45--54.

\bibitem{kye-2by2_II} S.-H. Kye, \it Facial
structures for unital positive linear maps in the two dimensional matrix
algebra, \rm Linear Alg. Appl., \bf 362 \rm (2003), 57--73.


\bibitem{kye-decom}  S.-H. Kye, \it Facial structures for
decomposable positive linear maps in matrix algebras, \rm
Positivity, to appear.

[avaible at {\tt http://www.math.snu.ac.kr/$\sim$kye/paper.html}]

\bibitem{lew00} M. Lewenstein, B. Kraus, J. I. Cirac and P.
Horodecki, \it Optimization of entanglement witness, \rm Phys.
Rev. A (5)\bf 62 \rm (2000), 052310

\bibitem{lew01} M. Lewenstein, B. Kraus, P. Horodecki and J. I.
Cirac, \it Characterization of separable states and entanglement
witnesses, \rm Phys. Rev. A (4) \bf 63 \rm (2001), 044304

\bibitem{stormer82}  E. St\o rmer, \it Decomposable positive maps on
$C^*$-algebras, \rm Proc. Amer. Math. Soc. \bf 86 \rm (1982),
402--404.

\bibitem{terhal00} B. M. Terhal and P. Horodecki, \it Schmidt
number for density matrices, \rm Phys. Rev. A (3) \bf 61 \rm
(2000), 040301

\bibitem{terhal01} B. M. Terhal, \it A family of indecomposable
positive linear maps based on entangled quantum states, \rm Linear
Algebra Appl. \bf 323 \rm (2001), 61--73



\bibitem{woronowicz}  S. L. Woronowicz, \it Positive maps of low
dimensional matrix algebras, \rm Rep. Math. Phys. \bf 10 \rm (1976),
165--183.


\end{thebibliography}
\end{document}